\documentclass{aa}

\usepackage{graphicx}
\usepackage{txfonts}
\usepackage{color}
\usepackage{url}
\usepackage[flushleft]{threeparttable}
\usepackage[colorlinks=true, allcolors=blue]{hyperref}
\usepackage{orcidlink}
\usepackage{adjustbox}
\usepackage{multicol}
\usepackage{multirow}
\usepackage{xspace}
\newcommand {\gx}{{GX~340$+$0}\xspace}
\newcommand {\ixpe}{{IXPE}\xspace}
\newcommand {\nustar}{\textit{NuSTAR}\xspace}
\newcommand {\maxi}{{MAXI}\xspace}
\newcommand {\xmm}{\textit{XMM--Newton}\xspace}
\newcommand {\chandra}{\textit{Chandra}/HETG\xspace}
\newcommand {\nicer}{{NICER}\xspace} 
\newcommand{\fluxcgs}{erg\,
s$^{-1}$\,cm$^{-2}$}

\begin{document}

\title{X-ray spectropolarimetric characterization of \gx in the horizontal branch: A highly inclined source?}

\titlerunning{X-ray spectropolarimetry of \gx}

\author{
Fabio La Monaca\inst{\ref{in:INAF-IAPS},\ref{in:UniRoma2},\ref{in:LaSapienza}}\thanks{Corresponding author: fabio.lamonaca@inaf.it}\orcidlink{0000-0001-8916-4156}
\and Alessandro Di Marco\inst{\ref{in:INAF-IAPS}}\orcidlink{0000-0003-0331-3259} 
\and
Renee M. Ludlam\inst{\ref{in:Wayne}}\orcidlink{0000-0002-8961-939X} 
\and  
Anna Bobrikova\inst{\ref{in:UTU}}\orcidlink{0009-0009-3183-9742}
\and 
Juri Poutanen \inst{\ref{in:UTU}}\orcidlink{0000-0002-0983-0049}
\and   \\  
Songwei Li \inst{\ref{in:Wayne}}\orcidlink{0009-0005-8520-0144}
\and
Fei Xie \inst{\ref{in:Guangxi},\ref{in:INAF-IAPS}}\orcidlink{0000-0002-0105-5826}
}

\authorrunning{F. La Monaca et al.}

\institute{
        INAF Istituto di Astrofisica e Planetologia Spaziali, Via del Fosso del Cavaliere 100, 00133 Roma, Italy \label{in:INAF-IAPS}
        \and
        Dipartimento di Fisica, Universit\`{a} degli Studi di Roma ``Tor Vergata'', Via della Ricerca Scientifica 1, 00133 Roma, Italy \label{in:UniRoma2} 
        \and 
        Dipartimento di Fisica, Universit\`{a} degli Studi di Roma ``La Sapienza'', Piazzale Aldo Moro 5, 00185 Roma, Italy \label{in:LaSapienza}
        \and
        Department of Physics and Astronomy, Wayne State University, 666 W. Hancock St., 48201 Detroit (MI), USA \label{in:Wayne} 
        \and Department of Physics and Astronomy, 20014 University of Turku, Finland \label{in:UTU}
        \and
        Guangxi Key Laboratory for Relativistic Astrophysics, School of Physical Science and Technology, Guangxi University, Nanning 530004, China \label{in:Guangxi}
        }
        
\date{Received 23 August 2024; accepted 2 October 2024}

\abstract{We report the first detection of X-ray polarization in the horizontal branch for \gx as obtained by the Imaging X-ray Polarimetry Explorer (\ixpe). A polarization degree of 4.3\%$\pm$0.4\% at a confidence level of 68\% is obtained. This value agrees with the previous polarization measurements of Z-sources in the horizontal branch. The spectropolarimetric analysis, performed using a broadband spectral model obtained by \nicer and \nustar quasi-simultaneous observations, allowed us to constrain the polarization for the soft and hard spectral components that are typical of these sources. 
The polarization angle for the two components differs by ${\sim}40\degr$. This result can be explained by a misalignment of the NS rotation axis with respect to the accretion disk axis. 
We compared the results with the polarization that is expected in different models. Theoretical expectations for the polarization of the disk and the Comptonized components favor a higher orbital inclination for \gx than 60\degr, as expected for Cyg-like sources. This is in contrast with the results we report for the reflection component based on the broadband spectrum.}

\keywords{accretion, accretion disk --
                polarization -- stars: low-mass --  stars: neutron -- stars: individual: GX 340+0 -- X-rays: binaries
               }

\maketitle
%

\section{Introduction}\label{sec:intro}

\mbox{GX~340+0} (4U 1642$-$45) is a weakly magnetized neutron star (WMNS) in a low-mass X-ray binary (LMXB). In these systems, the NS accretes matter from the donor stellar companion via Roche-lobe overflow and forms an accretion disk. The decelerating matter of the disk interacts with the surface of the NS to form a boundary layer (BL) that is coplanar to the disk \citep{Shakura88,Popham01}, but this matter can also spread up to high latitudes along the NS surface and forms a spreading layer \citep[SL;][]{inogamov1999,suleimanov2006}. 
Based on its X-ray color-color (CCD) and/or hardness-intensity diagram (HID),  \mbox{GX~340+0} is classified as a Z-source \citep[see, e.g.,][]{vanderklis89,hasinger89} and belongs to the so-called Cyg-like sources, together with \mbox{Cyg~X-2} and \mbox{GX~5$-$1} \citep{Kuulkers94, kuulkers97}. Their Z-track can be divided into different branches: horizontal (HB), normal (NB), and flaring (FB), separated by a hard apex (HA) and a soft apex (SA). Each branch corresponds to a different hardness of the source spectrum (softer for the FB and harder for the HB) with different accretion rates \citep{Church12, Motta19}. The other three known persistent Z-sources (\mbox{Sco~X-1}, \mbox{GX~349$+$2}, and \mbox{GX~17$+$2}) are classified as Sco-like sources, which show some differences from Cyg-like sources. The Sco-like sources are thought to have a lower inclination or a lower magnetic field \citep{kuulkers97, Psaltis95}. 
 
The spectra of WMNSs are dominated by a softer thermal emission from the disk or NS surface, and a harder component associated with a Comptonization due to the emission from the BL and/or SL. In addition, an iron line at energies above 6\,keV is usually present as the main reflection feature from the inner accretion disk. Gravitational redshift, Doppler shift, and relativistic Doppler boosting shape the iron line profile, and this allows us to investigate the environment near the compact object, which is dominated by strong gravity effects \citep[see, e.g.,][]{Cackett08, Cackett10, Cackett12, Matt06, Ludlam18,Ludlam22,  Mondal20,  Mondal22, Ludlam24}. The study of spectral properties in WMNSs, combined with the timing properties of the quasi-periodic oscillations  \citep{gilfanov2003,Revnivtsev06,Revnivtsev13}, allowed us to gain a deeper understanding of these systems, even though a full constraint of the geometry was not possible because the spectral and timing properties of the different emitting regions are degenerate. 

\begin{table}
\centering
\caption{List of telescopes/instruments with the respective observation IDs and the exposure times of each observation.}
\label{tab:exposure}
\begin{tabular}{llcc}
\hline \hline
& Obs ID &Telescope& Exp. time (s)\\
 \hline\

\ixpe & 03003301 & DU 1 & 100228 \\
& & DU 2 & 100234\\
& & DU 3 &  100157\\
\hline
\nicer &  7108010101 & & 1278 \\
\hline
\nustar & 91002313004 & FPMA & 12526 \\
& & FPMB &  12927\\
\hline
\end{tabular}
\end{table}

The launch of the Imaging X-ray Polarimetry Explorer \citep[\ixpe;][]{Soffitta2021,Weisskopf2022} in 2021 added X-ray polarimetry as a novel tool for investigating the geometry and accretion mechanism of these systems \citep{Lapidus85,suleimanov2006}. \ixpe observed several WMNSs, both Z- and atoll sources. For Cyg-like Z-sources, \ixpe measured a higher polarization degree (PD) in the HB of ${\sim}4\%$ \citep[see, e.g.,][for \mbox{GX~5$-$1}]{Fabiani24} compared to that in the NB or FB, which is $\leq$2\% \citep[see, e.g.,][for \mbox{Cyg~X-2}, and \mbox{Sco X-1}, respectively]{Farinelli23, LaMonaca24}. Moreover, the polarization angle (PA) measured by \ixpe for \mbox{Cyg~X-2} is aligned with the position angle of the radio jet \citep{Spencer13}, which is likely perpendicular to the accretion disk. On the other hand, for \mbox{Sco X-1}, \citet{LaMonaca24} reported a highly significant measure of X-ray polarization ($\sim$1\%) in the SA with a PA rotated by ${\sim}46\degr$ with respect to the direction of the radio jet \citep{Fomalont01, Fomalont01b}. Variation in the PA with the source state was also reported for the peculiar sources \mbox{Cir~X-1} \citep{Rankin24},  \mbox{XTE J1701$-$462} \citep{Cocchi23}, and \mbox{GX~13$+$1} \citep{Bobrikova24a,Bobrikova24b}. In particular, in the latter, a rotation of the PA by ${\sim}70\degr$ was reported, with a corresponding variation of the PD up to ${\sim}5\%$, and variations in the energy dependence of the polarization. \ixpe also measured the X-ray polarization of several atoll sources that show a lower average polarization with respect to the Z-sources, but a dependence of the PD on energy was observed, in particular, for \mbox{GX~9+9} \citep{Ursini23}, \mbox{4U~1820$-$303} \citep{DiMarco23b}, and \mbox{4U~1624$-$49} \citep{Saade24}, In these sources, energy-resolved polarization reaches higher values than in Z-sources, for instance, up to $\sim$10\% for \mbox{4U~1820$-$303} in the higher part of the \ixpe energy band. 

The bright and persistent WMNS \gx lies close to the Galactic plane. It was discovered through an Aerobee rocket campaign in 1969 \citep{Friedman67,Rappaport71}. It showed a full Z pattern in the CCD \citep{vanParadijs98, Kuulkers96} with an extreme flaring branch that is softer than the FB \citep{Penninx91,
Jonker98, Pahari24}. The exact inclination of the source is still a matter of debate. \citet{Kuulkers96}, reported that \gx has an intermediate inclination between the high-inclination systems \mbox{Cyg~X-2} and \mbox{GX~5$-$1} and the low-inclination Sco-like systems. This result is based on the variability properties and on the orientation of the HB in the CCD, and it also correlates it with the quasi-periodic oscillations. Deriving the inclination by the reflection spectral model of the asymmetric relativistic iron line using the high-resolution \xmm spectra, \citet{Dai09} constrained the inclination of the system at ${\sim}35\degr$. This value also agrees with the spectral analysis obtained by \chandra observations \citep{Miller16}. However, \citet{Seifina13} proposed a model with a Comptonized corona that puffs up while the source changes state from HB through the NB to the FB. In this model, the corona affects the visibility of the NS and causes the observed dipping FB, suggesting a high inclination for \gx. The distance of the source was estimated to be 11$\pm3$\,kpc \citep{Fender00}. The study of the radio emission of the source showed that the emission is variable with the Z branches of the source, with some evidence of a correlation with the X-ray emission \citep{Penninx93, Oosterbroek94}. No evidence of a spatially resolved radio jet was found in \gx.

\section{Observations}\label{sec:observations}

\subsection{\ixpe}\label{subsec:ixpe}

\ixpe observed \gx starting from 2024 March 23 at 13:56~UTC to 2024 March 25 at 13:25~UTC for a total exposure of about 100\,ks per each \ixpe detector unit (DU). The \ixpe observation ID and the exposure time for each DU are reported in Table~\ref{tab:exposure}. We analyzed the \ixpe data following the standard software and tools, as described in detail in Appendix~\ref{app:data_ixpe}.
Figure~\ref{fig:IXPE_LC}-\textit{top} shows the \maxi\footnote{\href{http://maxi.riken.jp/top/index.html}{http://maxi.riken.jp/top/index.html}} \citep{MAXI} light curve, which includes the time of the \ixpe observation. The \ixpe light curve and the corresponding hardness ratio (HR; given by the counting rate in 4--8\,keV divided by the one in the 2--4\,keV energy band) are reported in the same figure. The \ixpe HR does not show any variation along the observation.

\begin{figure}
   \centering
   \includegraphics[width=\columnwidth]{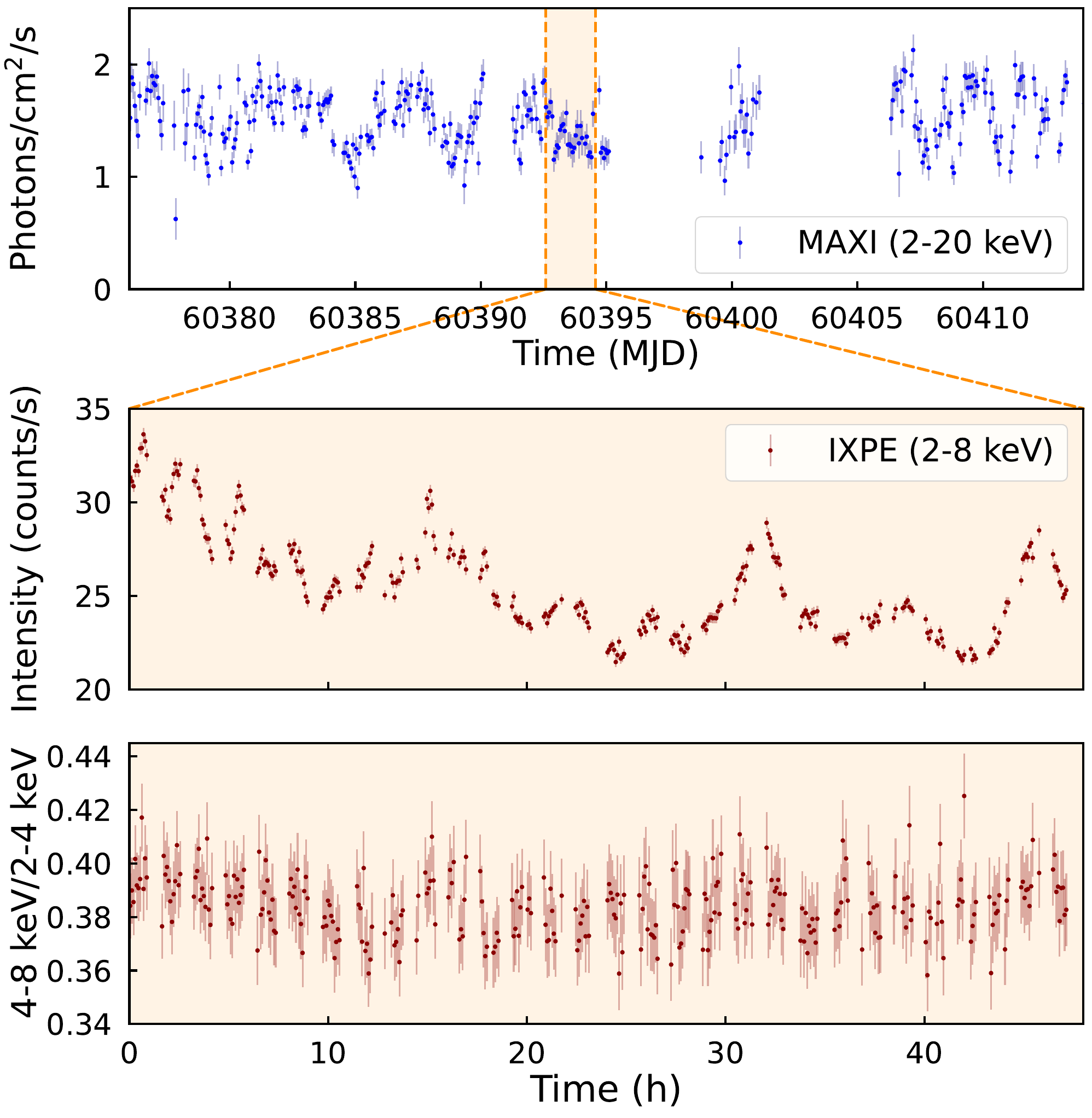}
   \caption{Light curves of \gx. Top panel: \maxi light curve binned into 1.5~h intervals, including the time of the \ixpe observation, which is highlighted in orange. Middle and bottom panels: \ixpe light curve with the corresponding HR binned in 300 s intervals. The HR is obtained as the ratio of the \ixpe counting rates in the 4--8 and 2--4\,keV energy bands.\label{fig:IXPE_LC}}
\end{figure}

\begin{figure}
   \centering
   \includegraphics[width=\columnwidth]{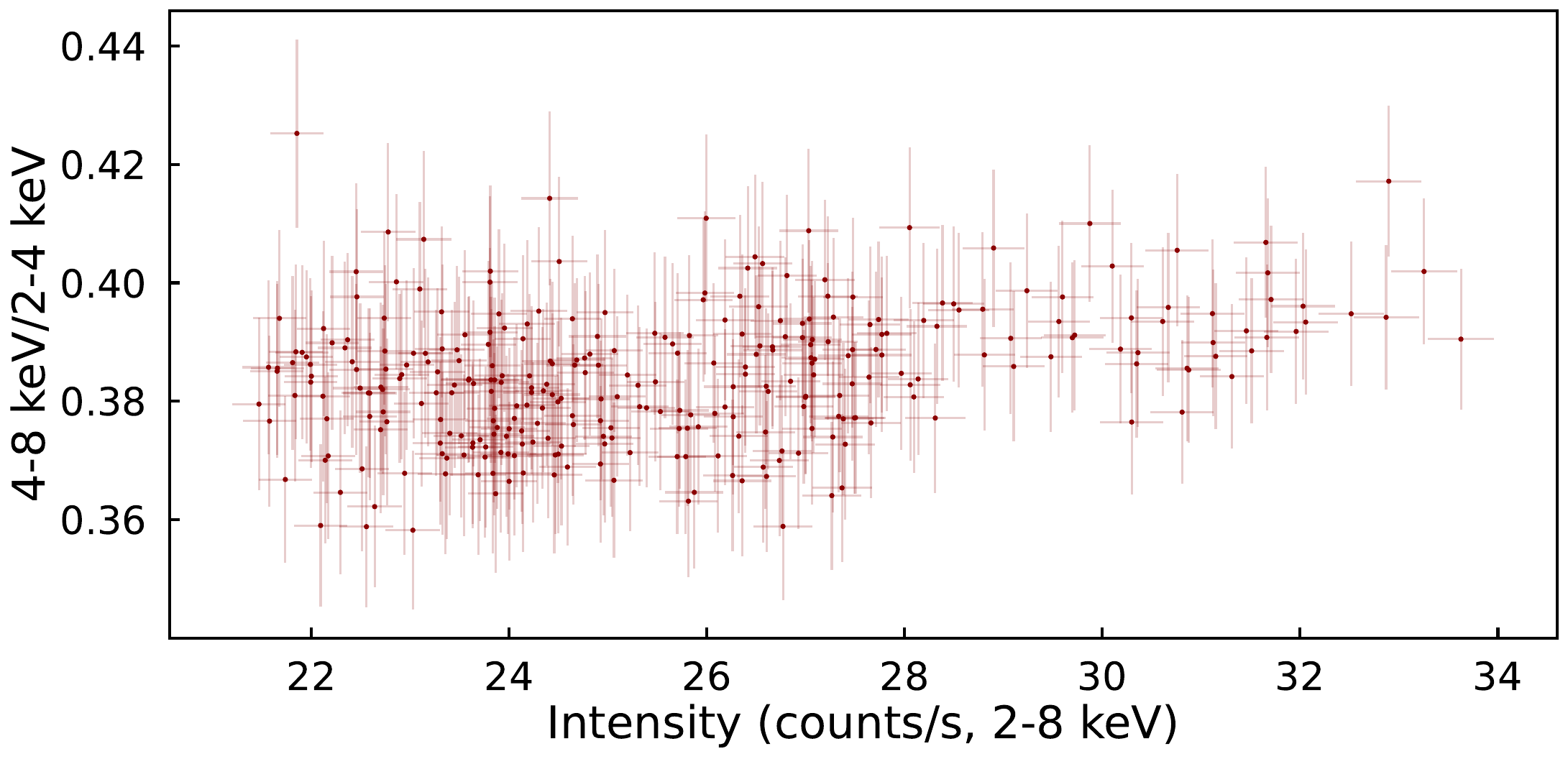}
   \caption{\ixpe Hardness-Intensity Diagram of \gx in 300\,s time bins. \label{fig:IXPE_HID}}
\end{figure}

\begin{figure} 
   \centering
   \includegraphics[width=0.88\columnwidth]{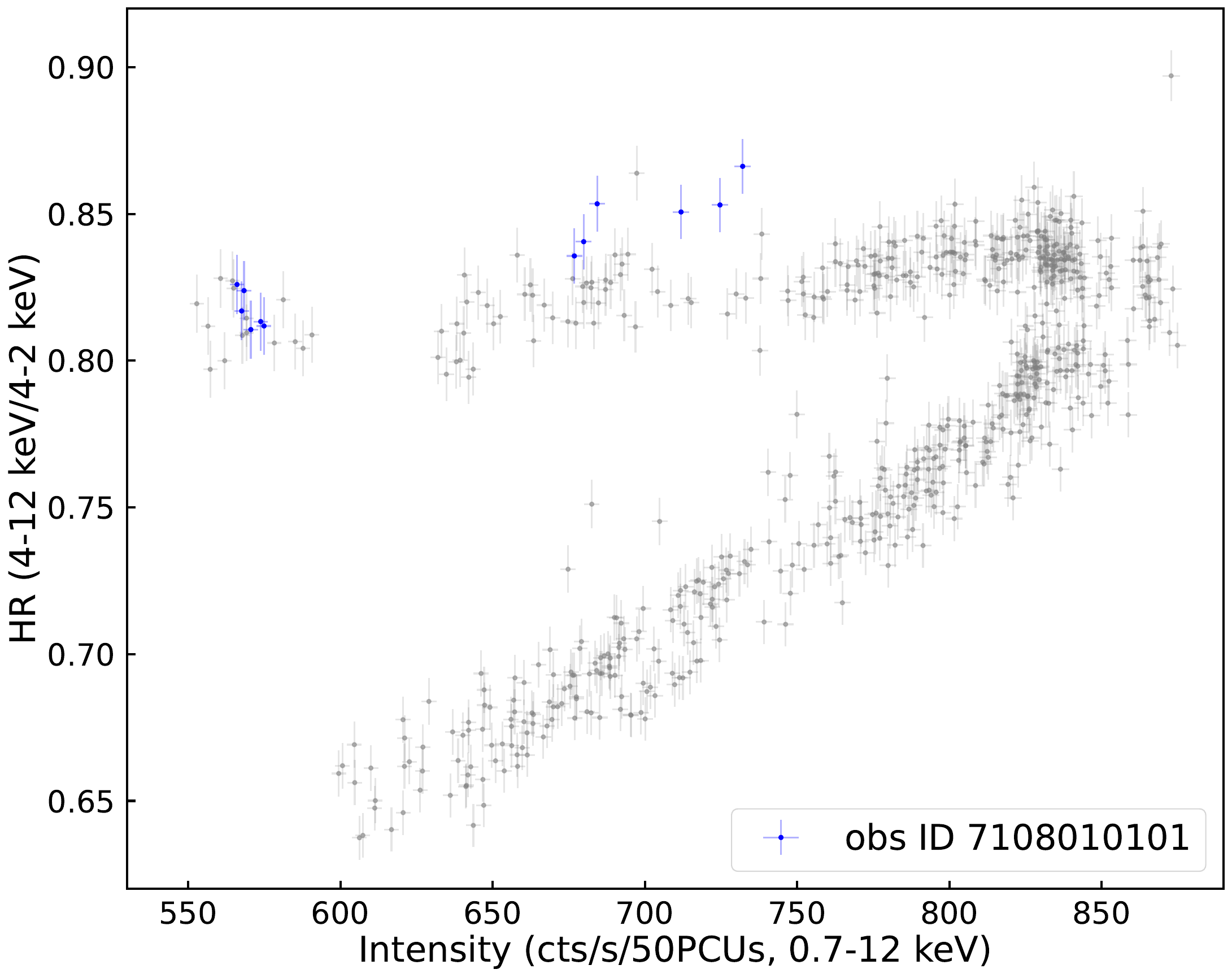}
   \includegraphics[width=0.88\columnwidth]{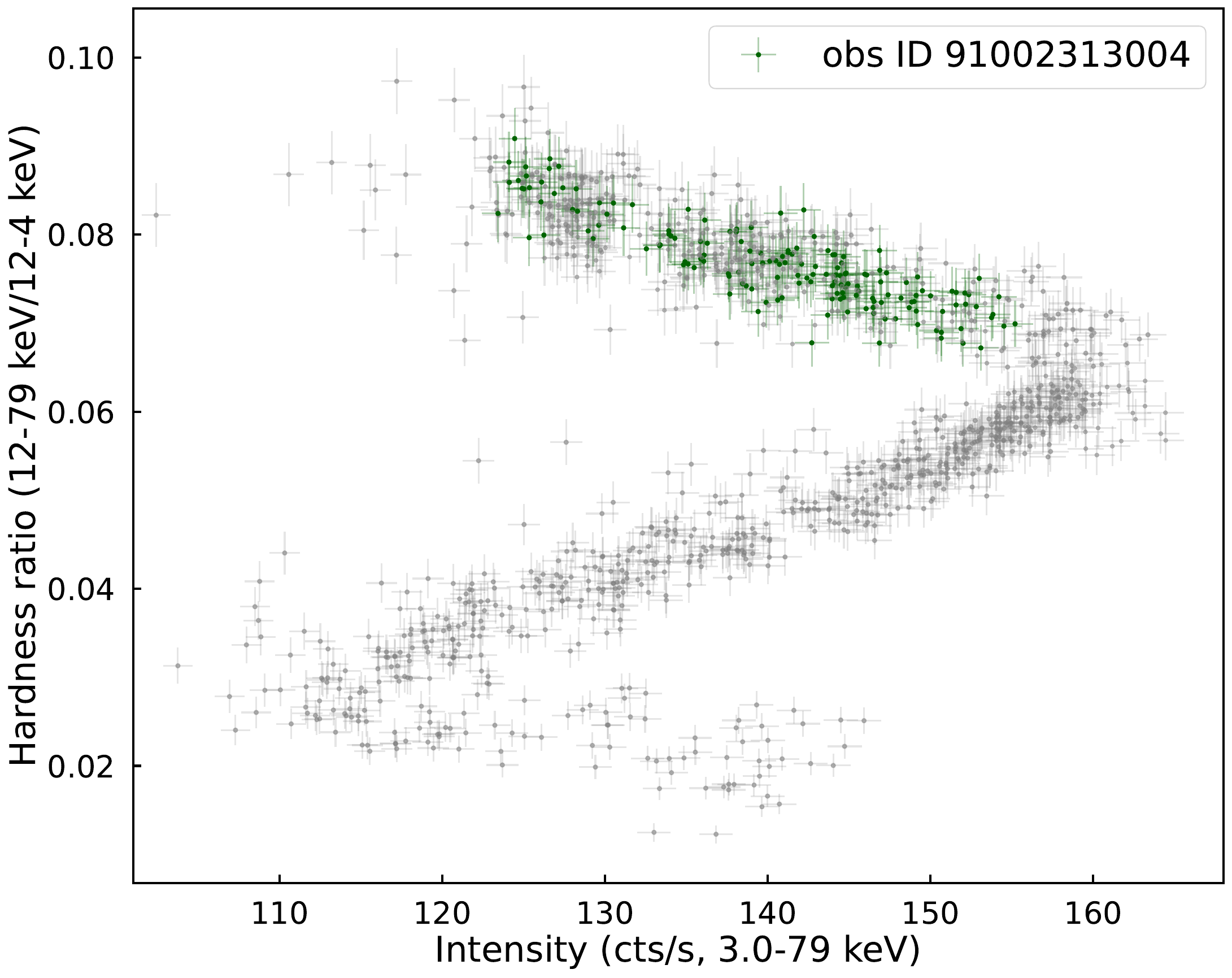} 
   \caption{Hardness-Intensity Diagram of \gx.  
   Top panel: HID of \gx from the \nicer data archive (gray points). The blue points represent the \nicer observation analyzed here. Bottom panel: HID of \gx  from the \nustar archival data  (gray points). The green points represent the \nustar observation analyzed here. In both the \nicer and \nustar observations analyzed here, \gx is in the HB, the same state as observed by \ixpe.\label{fig:NICER_NuSTAR_HID}}
\end{figure}

To determine the source state, we obtained the HID of the \ixpe observation shown in Fig.~\ref{fig:IXPE_HID}, where we plot the \ixpe hardness as a function of the intensity in the 2--8\,keV energy band. The hardness does not vary with intensity, and by combining this behavior with the absence of flaring activity in the light curve, we conclude that \gx was in the HB during the \ixpe observation. This confirms the result reported in \cite{Bhargava24}, who analyzed the same \ixpe observation.

\subsection{\nicer and \nustar}\label{subsec:NICER_NuSTAR}

Unfortunately, there are no strictly simultaneous observations with other X-ray observatories such as  \nicer or \nustar. They also cover different energy bands and might have helped identify the state of the source. The observations of \nicer and \nustar were performed three days before and two days after the \ixpe observation, respectively. To verify the state of the source during these observations, we compared the HID of these quasi-simultaneous observations with all the previous available observations of \gx in the \textsc{HEASARC} archive performed by \nicer and \nustar (see Fig.~\ref{fig:NICER_NuSTAR_HID}). The \nicer and \nustar data were extracted and analyzed using standard software and tools, as described in detail in the Appendices~\ref{app:data_nicer} and \ref{app:data_nustar}, respectively. Figure~\ref{fig:NICER_NuSTAR_HID} shows that during \nicer obs ID 7108010101 and \nustar obs ID 91002313004 (which were not simultaneous with the \ixpe observation, but had a gap of a few days), the source was in the HB. Therefore, we can use these two observations in addition to the observation obtained by \ixpe to model the spectrum.

\begin{figure}
   \centering
   \includegraphics[width=\columnwidth]{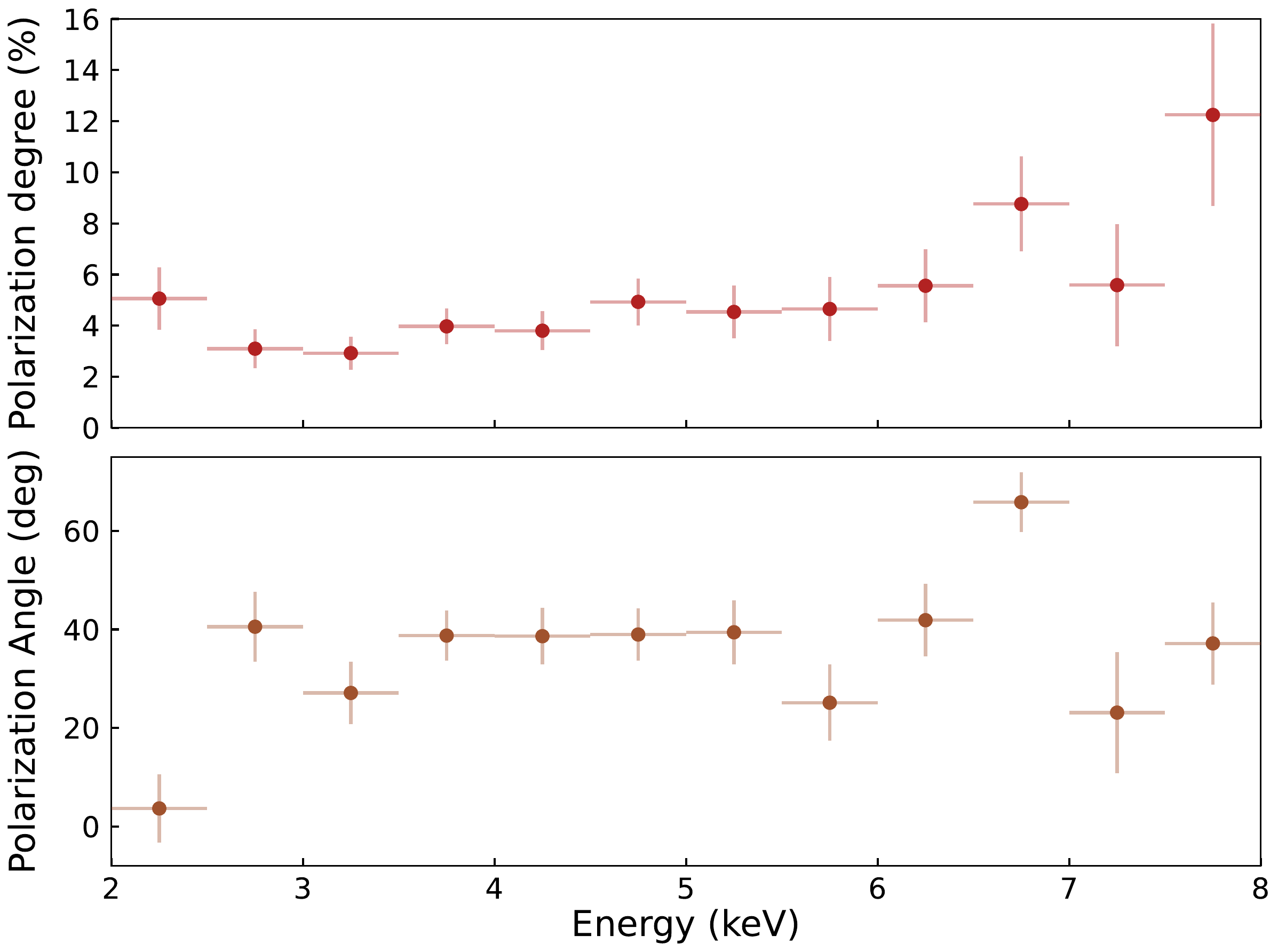}
   \caption{Energy-resolved PD (top panel) and PA (bottom panel) of \gx as a function of the energy in 0.5\,keV energy bins, as reported in Table~\ref{tab:PD_PAvsEnergy}. The errors are at 68\%~CL.\label{fig:PD_PAvsEnergy}}
\end{figure}

\begin{table}
\centering
\caption{Measured polarization for \gx.}
\label{tab:PD_PAvsEnergy}
\begin{tabular}{ccc}
\hline \hline
Energy Bin & PD & PA\\
(keV) & (\%) & (deg)\\
 \hline\
2.0--2.5 & $5.0\pm1.2$ & $4\pm7$  \\
2.5--3.0 & $3.1\pm0.8$ & $41\pm7$ \\
3.0--3.5 & $2.9\pm0.6$ & $27\pm6$ \\
3.5--4.0 & $4.0\pm0.7$ & $39\pm5$ \\
4.0--4.5 & $3.8\pm0.8$ & $39\pm6$ \\
4.5--5.0 & $4.9\pm0.9$ & $39\pm5$ \\
5.0--5.5 & $4.5\pm1.0$ & $39\pm6$ \\
5.5--6.0 & $4.6\pm1.3$ & $25\pm8$ \\
6.0--6.5 & $5.6\pm1.4$ & $42\pm7$ \\
6.5--7.0 & $8.8\pm1.1$ & $66\pm6$ \\
7.0--7.5 & $5.6\pm2.4$ & $23\pm12$ \\
7.5--8.0 & $12.2\pm3.6$ & $37\pm8$ \\
\hline
2.0--8.0 & 4.3$\pm$0.4 & 36$\pm$3 \\
\hline
\end{tabular}
\tablefoot{PD and PA values are obtained by the \texttt{pcube} algorithm in the \textsc{ixpeobssim} package in the entire 2--8\,keV energy band and in 0.5\,keV energy bins. The errors are reported at 68\%~CL.}
\end{table}

\section{Polarimetric analysis}

We estimated the polarization with a model-independent analysis (see Appendix \ref{app:data_ixpe}) using the \texttt{pcube} algorithm in the \textsc{ixpeobssim} software \citep{Baldini22}. The measured X-ray polarization in the nominal 2--8\,keV energy band for \gx is $\textrm{PD} = 4.3\%\pm0.4\%$, and the $\textrm{PA} = 36\degr\pm3\degr$ at a confidence level (CL) of 68\%. The significance of the detection is extremely high, above $\sim$11$\sigma$ CL. This means that the probability of obtaining this polarization in the case of an unpolarized source is $2\times 10^{-27}$.

Because the detection is highly significant, we can divide the \ixpe energy band into 0.5\,keV energy bins. The measured polarization in each energy bin exceeds 99\%~CL (see Fig.~\ref{fig:PD_PAvsEnergy}), except for the bin in the range 7.0--7.5\,keV, which has a significance at 93\%~CL. The measured values at 68\%~CL in each energy bin are reported in Table~\ref{tab:PD_PAvsEnergy}. It should be noted that the PA in the first energy bin (2.0--2.5\,keV) differs from the following bins by ${\sim}37\degr$ at 96\%~CL, and a slight variation in the PA is also visible for the 6.5--7.0\,keV energy bin. This last bin also indicates an increase in the PD. All other energy bins have an average PA of $\sim$35\degr. Moreover, the PD in the last bin (7.5--8.0\,keV) is $12.2\%\pm3.6\%$ at a $\textrm{PA}=37\degr\pm8\degr$ that is significant at 99.7\%~CL. This value is compatible with the value measured for \mbox{4U 1820$-$303}, where a PD of $10.3\%\pm2.4\%$ was measured in the 7--8\,keV energy bin \citep{DiMarco23b}.

\begin{figure}
   \centering
   \includegraphics[width=\columnwidth]{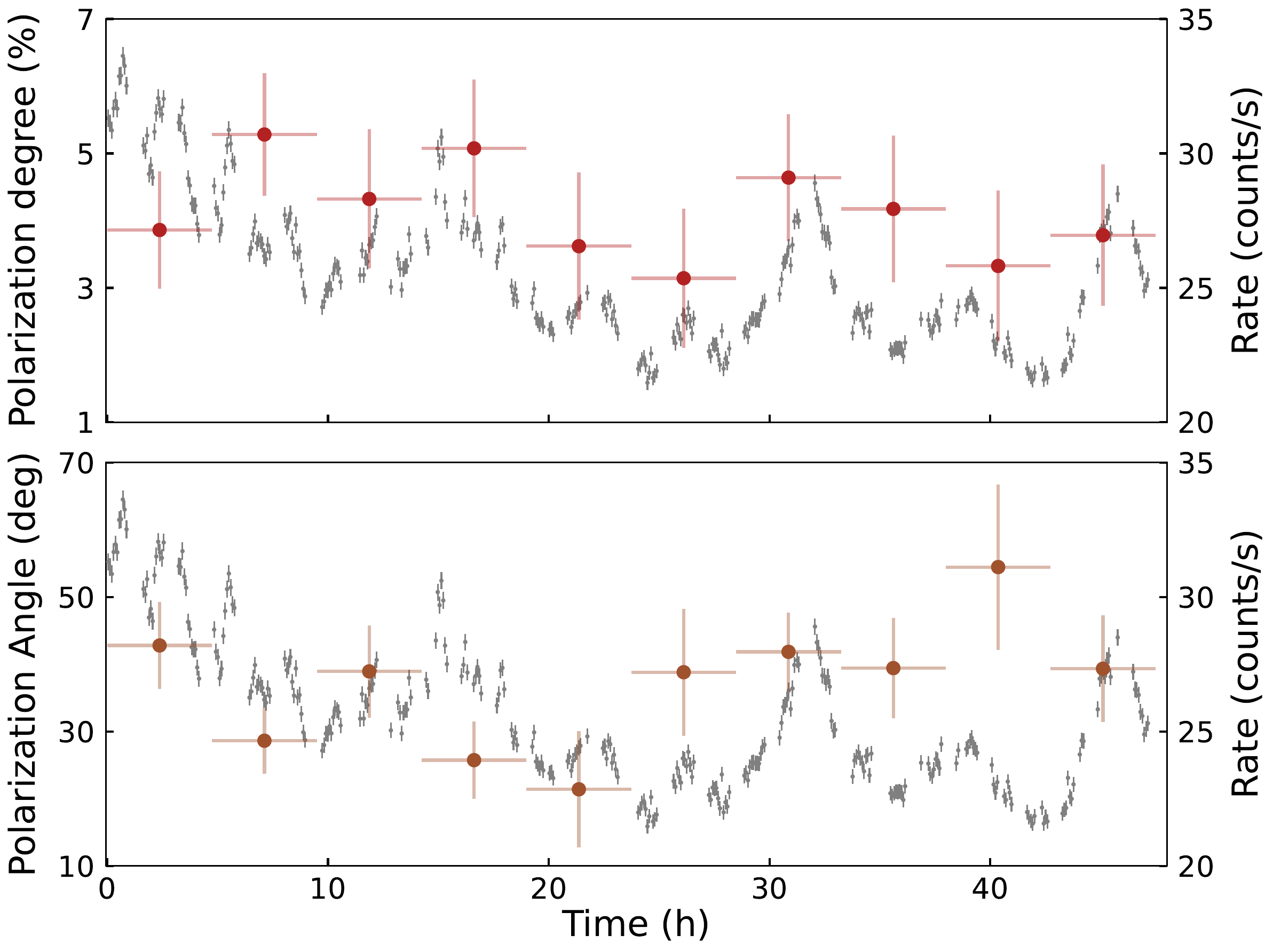}
   \caption{Time dependence of the PD (top panel) and PA (bottom panel) of \gx. The observation is divided into ten equal time bins of ${\sim}4.7$~hours. The errors are reported at the 68\%~CL \label{fig:PD_PAvsTime}}
\end{figure}

We also performed a time-resolved polarimetric analysis because we have evidence of polarization variability with time in some WMNSs, for example, in \mbox{GX 13$+$1} \citep{Bobrikova24a,Bobrikova24b}. Figure~\ref{fig:PD_PAvsTime} shows the resulting PD and PA in ten equally spaced time bins of ${\sim}4.7$~h. The results are compatible at the 68\%~CL, reporting a constant behavior in time and/or with the flux of the source. 

\section{Spectropolarimetric analysis} \label{sec:spec}

The \gx spectrum has been modeled using NICER (1.5--10\,keV), \nustar (3--20\,keV), and \ixpe (2--8\,keV) weighted spectra \citep{DiMarco_2022}. We first used the \texttt{tbabs*(diskbb+comptt+gauss)} model from \textsc{xspec}, noted as spectral Model~A in Table~\ref{tab:spectrum}. To estimate the absorption from the interstellar medium, we set the abundances at the \texttt{wilm} values in \textsc{xspec}, which use the values reported in \cite{2000ApJ...542..914W}. The \texttt{diskbb} model was used to describe the soft spectral component, and the \texttt{comptt} model represented the harder spectral component in which photons are up-scattered via Comptonization. The Gaussian line was used to model an excess at ${\sim}6.7$\,keV related to a Fe line expected in the presence of reflection from the disk. In the \nicer spectrum, we included an edge at 1.84\,keV and a line at 1.74\,keV to compensate for instrumental effects related to Si in the windows and bulk detectors. The parameters we obtained for this spectral model are reported in Table~\ref{tab:spectrum}. To take calibration uncertainties into account \citep{madsen22, DiMarco_2022b}, we kept a gain offset free in the fit for \nustar and \ixpe and obtained a slope with a deviation from one below the uncertainties of the fit and a gain offset smaller than one bin. A cross-normalization constant for each detector was applied, as reported in Table~\ref{tab:spectrum}.

\begin{figure}
   \centering
   \includegraphics[width=\columnwidth]{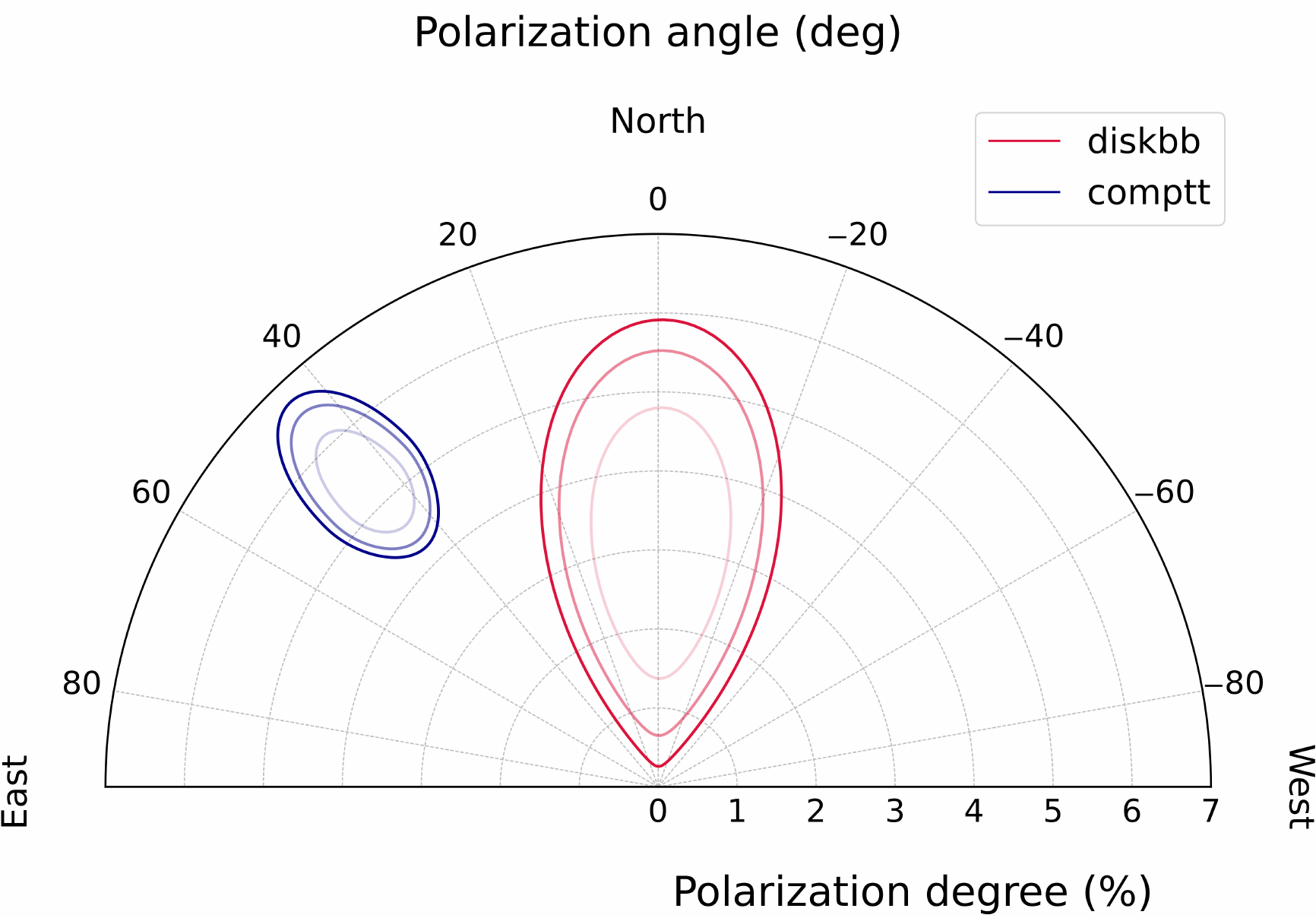}
   \caption{Polar plot with the results of the spectropolarimetric analysis, in which a \texttt{polconst} component is associated with the \texttt{diskbb} and \texttt{comptt} components separately, without any prior constraints on the PA. The contours correspond to 68\%, 90\%, and 95.5\%~CL from the innermost to the outermost edge. \label{fig:contours_spectropol}}
\end{figure}

The \ixpe Stokes parameters $I$, $Q$, and $U$ spectra were fit by freezing the spectral model to the best-fit parameters for Model~A, reported in Table~\ref{tab:spectrum}. The $I$ spectrum alone gives a $\chi^2/\textrm{d.o.f.} = 88/77 = 1.1$. The first spectropolarimetric analysis, including $Q$ and $U$ spectra, was performed to confirm the model-independent result for the polarization in 2--8\,keV energy band. We assumed a constant polarization using the \texttt{polconst} model from \textsc{xspec} for both the \texttt{diskbb} and \texttt{comptt} components, and the Gaussian was assumed to be unpolarized \citep{Churazov02}. We obtained the $\textrm{PD}=3.6\%\pm0.4\%$ at $\textrm{PA}=35\degr\pm3\degr$ with $\chi^2/\textrm{d.o.f.} = 257/249 = 1.0$; hereafter, errors are reported at the 90\%~CL. We also attempted to use the linear polarization model \texttt{pollin} from \textsc{xspec}, but the slope parameters for PD and PA remained undetermined. At this point, we tried to disentangle the polarization of the \texttt{diskbb} and \texttt{comptt} components by associating each component with its own polarization. We assumed a constant polarization for each of the components, obtaining for the \texttt{diskbb} a PD of $3.1\% \pm 1.7\%$ with a $\textrm{PA} = -1\degr \pm 16\degr$ and for \texttt{comptt} $\textrm{PD} = 5.2\% \pm 1.0\%$ with $\textrm{PA} = 44\degr \pm 5\degr$ and the best-fit $\chi^2/\textrm{d.o.f.}=242/247=0.98$. The polar plot representing the results of this spectropolarimetric analysis is shown in Fig.~\ref{fig:contours_spectropol}. This is one of the best estimates obtained from spectropolarimetric analysis of \ixpe data so far of the polarization for the \texttt{diskbb} and Comptonized components when both were left free to vary without any constraints on the PA. 

\begin{table*}
\centering
\caption{Best-fit parameters of the spectral models applied to the joint data from \nicer, \nustar, and \ixpe.}
\label{tab:spectrum}
\begin{tabular}{lrcc}
\hline \hline
Model & Parameter (units) & Model A & Model B \\ \hline
\texttt{TBabs} & $N_{\rm H}$ ($10^{22}$ cm$^{-2}$) & $8.58_{-0.08}^{+0.09}$ & $8.70_{-0.05}^{+0.03}$ \\ \hline
\texttt{diskbb} & $kT_{\rm in}$ (keV) & $0.87\pm0.06$ & $1.36_{-0.02}^{+0.04}$\\
                & norm ($[R_{\rm in}/D_{10}]^2\cos\theta$) & $670_{-110}^{+280}$ & $166_{-15}^{+7}$\\ \hline
\texttt{comptt} & $T_0$ (keV) & $1.19\pm0.03$ & $1.90_{-0.07}^{+0.06}$\\
                & $kT$ (keV) & $3.5_{-0.10}^{+0.06}$ & $3.3652_{-0.0011}^{+0.0004}$\\
                & $\tau$   & $4.8_{-0.05}^{+0.17}$ & $3.7_{-0.2}^{+0.9}$\\
                & norm     & $0.70\pm0.02$ & $0.278_{-0.006}^{+0.040}$\\ \hline
\texttt{Gaussian} & $E$ (keV) & $6.65_{-0.06}^{+0.05}$ & --\\
                & $\sigma$ (keV) & $0.50_{-0.10}^{+0.06}$ & --\\
                & norm (photon~cm$^{-2}$~s$^{-1}$) & $0.0043\pm0.0005$ & --\\
                & Equivalent width (eV) & $38\pm8$ & --\\ \hline
\texttt{relxillNS} & Emissivity & -- & [3.0] \\
                & $R_{\rm in}$ (\texttt{ISCO}) & -- & [1.44] \\
                & $R_{\rm out}$ ($GM/c^2$) & -- & [1000] \\
                & Inclination (deg) & -- & $36.7_{-1.4}^{+2.9}$\\
                & $\log \xi$ & -- & $2.72_{-0.14}^{+0.10}$\\
                & $A_{\rm Fe}$ & -- & $0.70_{-0.19}^{+0.33}$\\
                & $\log N$ & -- & [19] \\
                & \text{norm} & -- & $0.0032_{-0.0006}^{+0.0010}$\\
\hline
\multicolumn{2}{r}{$\chi^2/\textrm{d.o.f.}$} & 1470/1394 = 1.05 & 1535/1406 = 1.09\\ \hline
\multicolumn{4}{c}{Cross normalization factors} \\
&  $C_{\rm NICER}$ & 1.0 & 1.0\\
& $C_{\rm \nustar-A}$ & $1.15\pm0.02$ & $1.031\pm0.004$\\
& $C_{\rm \nustar-B}$ & $1.24\pm0.02$  & $1.110_{-0.010}^{+0.002}$\\
&  $C_{\rm \ixpe-DU1}$ & $0.749\pm0.015$ & $0.699\pm0.003$\\
& $C_{\rm \ixpe-DU2}$ & $0.760\pm0.015$ & $0.708\pm0.003$\\
& $C_{\rm \ixpe-DU3}$ & $0.749\pm0.015$ & $0.697_{-0.004}^{+0.002}$\\ \hline
\multicolumn{4}{c}{Photon flux ratios in 2--8\,keV} \\
& $F_{\rm \texttt{diskbb}}/F_{\rm tot}$ & 0.210 & 0.540\\
& $F_{\rm \texttt{comptt}}/F_{\rm tot}$ & 0.784 & 0.337\\
& $F_{\rm \texttt{gauss}, \texttt{relxillNS}}/F_{\rm tot}$ & 0.006 & 0.123\\ \hline
\end{tabular}
\tablefoot{The estimated flux in 2--8\,keV is $5\times10^{-9}$\,\fluxcgs. The errors are reported at 90\%~CL.}
\end{table*}

As a further step, we tried another model, where the reflection component is described by the \texttt{relxillNS} model \citep{Garcia22}: \texttt{tbabs*(diskbb+comptt+relxillNS)}, noted as Model~B in Table~\ref{tab:spectrum}. We assumed that the X-ray continuum is described by a blackbody originating from the NS surface or the BL and/or SL, with the same temperature as the \texttt{comptt} seed photons. This thermal component illuminates the cold accretion disk, producing reprocessed/reflected X-rays of the reflection component. The density $\log N$ parameter was frozen to the maximum value allowed for the model \citep{Garcia22}, and the outer radius was fixed to $1000\,R_{\rm g}$. The reflection fraction parameter was fixed to $-1$, and we therefore selected only the spectral component due to reflection, while the emissivity was frozen to $q=3$ as in \cite{2018MNRAS.475..748W}, and the disk inner radius $R_{\rm in}$ to 1.44 in units of the innermost stable circular orbit (ISCO) radius, as suggested for the same state and epoch of \gx by S. Li et al. (in prep.). The best-fit parameters are reported in Table~\ref{tab:spectrum} with errors at the 90\%~CL obtained from a Markov Chain Monte Carlo (MCMC) with 60 walkers, a burn-in of $1\times10^5$, and a chain length of $5\times 10^4$. When we associate a different constant polarization to each spectral component and apply the spectropolarimetric analysis to this model, we obtain the following upper limits ($\chi^2/\textrm{d.o.f.}=270/254=1.1$): $\textrm{PD}<7\%$ for the \texttt{diskbb}, $<40\%$ for the \texttt{comptt}, and $<72\%$ for the \texttt{relxillNS}. Upon fixing the PD of the \texttt{comptt} at 0\%, we obtain for the disk $\textrm{PD}=2.6\% \pm1.4\%$ with $\textrm{PA}=8\degr \pm16\degr$, and for the reflection $\textrm{PD}=29\% \pm 9\%$ with $\textrm{PA}=49\degr \pm 9\degr$ ($\chi^2/\textrm{d.o.f.}=270/256=1.05$). When we fixed the polarization degree for Comptonization to higher values, the reflection polarization decreased. The higher the \texttt{comptt} polarization, the smaller the polarization associated with the reflection component. This confirms the difficulties of disentangling the polarization for these two components. 

\section{Discussion and conclusions}

\gx was observed by \ixpe when the source was in the HB (see Fig.~\ref{fig:IXPE_HID}). \ixpe measured a  highly significant average polarization ($\sim$11$\sigma$~CL) with the $\textrm{PD}=4.3\%\pm0.4\%$ at $\textrm{PA}=36\degr\pm3\degr$ in the 2--8\,keV energy band using the model-independent \texttt{pcube} analysis. These results agree with those obtained by the spectropolarimetric analysis.

Before \gx, \ixpe observed two Z-sources in the HB: \mbox{GX 5$-$1} and \mbox{XTE J1701$-$462}. For the latter, a $\textrm{PD}=4.6\%\pm0.4\%$ with $\textrm{PA}=-38\degr\pm2\degr$ was observed, with no rotation of the PA or variation in PD over time, but with an indication of a variation in PD in the 2--3\,keV energy bin with respect to higher energies \citep[see Fig.~8 of][]{Cocchi23}. For \mbox{GX 5$-$1}, an average PD of $4.3\%\pm0.3\%$ with a PA of $-10\degr\pm2\degr$ is measured in the HB; no variation in PD with time or energy were reported, while a possible PA variation with energy was claimed. Therefore, we can conclude that the average PD of the HB of \gx agrees very well with the other previous measurements of the WMNSs in the same state. The Z-sources in the HB show a higher PD ($\sim$4\%) than the other branches, where the measured PD is $\sim$1\%. \ixpe observed two other Z-sources, but not in the HB: \mbox{Cyg~X-2} in NB with a measured polarization $1.8\% \pm 0.3\%$ and PA of 140\degr$\pm$4\degr, aligned with previous measurements of the radio jet position angle, and \mbox{Sco~X-1} in SA, measuring a PD of $1.0\% \pm 0.2\%$ with a PA of 8\degr$\pm$6\degr\ at the 90\%~CL, not aligned with the radio jet position angle. \cite{Oosterbroek94} reported a correlation between the \gx radio flux and the X-ray spectral state, similar to the one reported for other Z-sources \citep[see, e.g.,][]{Migliari2006}, but the radio jet was not spatially resolved, and thus, a comparison with the PA is not possible for \gx.  

\gx shows a PD and PA trend with energy similar to the results obtained for \mbox{GX 5$-$1} and \mbox{XTE J1701$-$462}. Because the detection was highly significant, we were able to perform an energy-resolved polarization analysis, for which we divided the nominal \ixpe energy band into 0.5\,keV energy bins. This analysis showed interesting results at both higher and lower energies. The lower-energy bin is rotated with respect to the next energy bin by ${\sim}37\degr$ at the 95\%~CL. In aligned systems, a 90$\degr$ rotation is expected for the soft component compared to the Comptonized component, but the measured lower rotation is somewhat unexpected and might be related to a misalignment of the NS rotation axis from the accretion disk axis \citep{Bobrikova24b,Rankin24}. Moreover, the highest-energy bin shows an extremely high polarization of $12.2\%\pm3.6\%$ with a $\textrm{PA}=37\degr\pm8\degr$ that is significant at the 99.97\%~CL. This value is compatible with the highest measured value by \ixpe in WMNSs, where a PD of $10.3\%\pm2.4\%$ was measured in the 7--8\,keV energy bin in the ultracompact atoll source \mbox{4U~1820$-$303} \citep{DiMarco23b}. A high PD like this is not expected by standard corona geometries. 

We performed a spectral analysis of \gx in HB by combining the \ixpe observation with those by \nicer and \nustar in the same state, as reported in Sect.~\ref{sec:spec}. The estimated fluxes showed a spectrum that is dominated by the Comptonized component (${\sim}78\%$ of the total flux), with the soft component contributing to the total flux with $\sim$21\%  and the Gaussian line only with $\sim$0.6\%, with an equivalent width of $38 \pm 8$\,eV. Using this simple spectral model (noted as Model~A in Table~\ref{tab:spectrum}), we performed a spectropolarimetric analysis and obtained an uncommon PA difference of ${\sim}40\degr$ between the two components. This confirms the result reported in Table~\ref{tab:PD_PAvsEnergy}, which was previously discussed, of the rotation of PA in the lowest-energy bin (almost aligned with the \texttt{diskbb}) with respect to the following ones, which have an average $\textrm{PD}{\sim}5\%$ and $\textrm{PA}{\sim}40\degr$  that is consistent with the polarimetric result for the \texttt{comptt} component.
This difference in PA is similar to the difference reported in \citet{Bobrikova24b} for GX~13+1. For aligned systems, the theoretical models  \citep[such as, e.g.,][]{st85} predict that the PA of the disk and the Comptonized components are almost parallel or orthogonal, depending on the optical depth of the Comptonizing media. As suggested by \citet{Bobrikova24b} and \citet{Rankin24}, to obtain a difference in the PA such as we report here for \gx, some misalignment of the disk/BL axis with the NS axis is needed.

In the case of an optically thick accretion disk that is dominated by electron scattering, the PD of the disk component is expected to be $\sim$0.7\% when the inclination is 35\degr\ and $\sim$2\% when the inclination is 60\degr \citep{Chandrasekhar60, Sobolev63,Loktev22}. Thus, the result obtained here for the PD of ${\sim}3\%$ favors an inclination higher than 60\degr, as expected for Cyg-like sources and in \citet{Seifina13} and \citet{Kuulkers96}, rather than the 35\degr\  reported by \citet{Dai09} and \citet{Miller16}. For the Comptonized component, considering the results reported in Fig.~5 of \cite{st85}, which was obtained for a slab with a Thomson optical depth of $\tau_{\rm T}$$\sim$4--5 and $kT_{\rm e}$$\sim$3\,keV, the measured PD of ${\sim}5\%$ is not compatible with an inclination of ${\sim}35\degr$, but requires higher inclinations (${>}70\degr$). A similar result is also expected for an equatorial belt SL, as reported by \cite{Farinelli24, Bobrikova24c}. However, we note that in the case of the presence of wind, which is indicated for \gx in \cite{Miller16}, higher values of the PD can also be expected for lower inclinations, as reported by \cite{Tomaru2024}. Furthermore, as reported in Table~\ref{tab:PD_PAvsEnergy} and confirmed by spectropolarimetric analysis, the PD is almost constant above 2.5\,keV with an increase at higher energies. \cite{Poutanen_2023} proposed a model for the black hole X-ray binary Cyg X-1 in the hard state, in which the hot medium is situated within a truncated cold accretion disk, with the seed unpolarized photons coming from the outer cold truncated disk~(case C) or produced within the flow~(case B). The observed behavior of polarization for \gx might be compatible with a corona geometry like this. For a system inclination of 60\degr, the expected PD in both cases is higher than the value measured here for the Comptonized component, but when the inclination is 30\degr, the expected PD is too small. An intermediate inclination of the system might therefore be compatible with our results. However, the analysis of the reflection spectrum points to an inclination lower than 40\degr. We also attempted to fit the \gx spectrum with a model that included a reflection component, noted as Model~B in Table~\ref{tab:spectrum}. From this analysis, we obtained an inclination of ${\sim}37\degr$, which is compatible with the results of \cite{Dai09}, \cite{Miller16}, and Li et al. (in prep.). We attempted to determine the polarization for this reflection component and found that this value depends strongly on our assumptions for the polarization of the Comptonized component. When we left the PD for each component in Model~B free to vary, we only obtained upper limits for each of them. When we assumed that the \texttt{comptt} spectral component was unpolarized, almost the case of \cite{st85} and \cite{Farinelli24} for an inclination of ${\sim}35\degr-40\degr$, the reflection component reached a PD up to ${\sim}30\%$. This is in line with the results from theoretical expectations \citep{Matt93,Poutanen96}.
When the PD attributed to \texttt{comptt} was assumed to be higher, the PD of the reflection decreased. This confirms the degeneracy of the polarization of these two spectral components in our \ixpe data. Moreover, the reduced $\chi^2$ shows no preference for the spectropolarimetric fit with \texttt{relxillNS}, which is not capable of fully explaining the high PD value observed in the highest-energy bin, 7.5--8.0~keV, in our data set.

Future observations of \gx will help us to better characterize the polarimetric properties of the accretion flow. During this observation, \gx was in the HB. In the future, we hope to measure the polarimetric properties of \gx in different branches. The possible misalignment between the disk or BL components and/or the SL or NS axis also requires further observations to confirm or discard the indication.

\begin{acknowledgements}
This research used data products provided by the IXPE Team (MSFC, SSDC, INAF, and INFN) and distributed with additional software tools by the High-Energy Astrophysics Science Archive Research Center (HEASARC), at NASA Goddard Space Flight Center (GSFC). The Imaging X-ray Polarimetry Explorer (IXPE) is a joint US and Italian mission.  
The authors acknowledge the NuSTAR team for scheduling the observations. In particular, the authors acknowledge Karl Forster, Murray Brightman, Hannah Penn Earnshaw, Hiromasa Miyasaka, Daniel Stern, and Fiona A. Harrison. 
The Italian contribution is supported by the Italian Space Agency (Agenzia Spaziale Italiana, ASI) through contract ASI-OHBI-2022-13-I.0, agreements ASI-INAF-2022-19-HH.0 and ASI-INFN-2017.13-H0, and its Space Science Data Center (SSDC) with agreements ASI-INAF-2022-14-HH.0 and ASI-INFN 2021-43-HH.0, and by the Istituto Nazionale di Astrofisica (INAF) and the Istituto Nazionale di Fisica Nucleare (INFN) in Italy. FLM and ADM are partially supported by MAECI with grant CN24GR08 “GRBAXP: Guangxi-Rome Bilateral Agreement for X-ray Polarimetry in Astrophysics”.
RML and SL acknowledge support by NASA under grant No. 80NSSC23K0498.
AB is supported by the Finnish Cultural Foundation grant No. 00240328.
FX is supported by the National Natural Science Foundation of China (Grant No. 12373041), and special funding for Guangxi distinguished professors (Bagui Xuezhe).

\end{acknowledgements}

%
%
\bibliographystyle{yahapj}
\bibliography{biblio}
-------------------------------------------------------------

\begin{appendix}
\section{Data Handling}\label{app:data}

The light curves and HID/CCD for the X-ray
observatories studied in this paper were obtained using \textsc{stingray}, a library developed in \textsc{Python} for spectral timing analysis \citep{stingray2,stingray1,Bachetti24}. In the following subsections, each X-ray observatory is discussed in more detail.

\subsection{\ixpe}\label{app:data_ixpe}

\ixpe is the first X-ray mission entirely dedicated to studying linear X-ray polarization thanks to three identical telescopes coupling a multi-mirror array and a detector unit (DU) hosting a photoelectric polarimeter. A description of the satellite's performances and the onboard photoelectric polarimeters can be found in \cite{Weisskopf2022}, \cite{Soffitta2021}, \cite{DiMarco22b}, \cite{Baldini21}, and references therein. We selected the source in a circular centered region of 100\arcsec\ radius, using \textsc{SAOImageDS9}\footnote{\href{https://sites.google.com/cfa.harvard.edu/saoimageds9}{https://sites.google.com/cfa.harvard.edu/saoimageds9}}, and following the \ixpe prescription reported in \cite{DiMarco23a} without subtracting the background due to the brightness of the source. The \ixpe data were processed with the \textsc{ixpeobssim} package version 31.0.1 \citep{Baldini22}, a software for polarimetric analysis developed for the \ixpe mission that includes model-independent analysis tools, using the \texttt{pcube} algorithm based on \citet{Kislat15}. Spectral and spectro-polarimetric analysis were performed extracting data using \textsc{ftools} included in HEASoft version 6.33.2 \citep{heasoft}, and using \textsc{xspec} \citep{Arnaud96} with \ixpe CALDB response matrices 20240101 released on  2024 February 28. For the spectral analysis, we binned the \ixpe data to have a minimum of 30 counts per bin, while for the spectro-polarimetric analysis, a constant binning of 200 eV has been applied.

\subsection{\nicer}\label{app:data_nicer}
\nicer is a non-imaging soft X-ray instrument operating in the 0.2--12\,keV on board the International Space Station; it consists of 56 silicon drift detectors placed at the focus of co-aligned concentrator X-ray optics \citep{Gendreau16}. We processed \nicer data using the standard pipeline provided by the \nicer team and analyzed with \nicer Data Analysis Software v012a, distributed in HEASoft v6.33.2 with the CALDB version \textsc{XTI20240206}. The \nicer spectra were grouped to have a minimum of 30 counts per bin, and the background spectra were obtained using the \nicer \textsc{SCORPEON} model. 

\subsection{\nustar}\label{app:data_nustar}

\nustar is an X-ray observatory consisting of two identical telescopes (FPMA and FPMB) \citep{Harrison2013}, each coupling a depth-graded multilayer-coated Wolter-I conical approximation X-ray optics and a solid-state focal plane CdZnTe pixel detector surrounded by a CsI anti-coincidence shield; it operates in the energy range 3--79\,keV, providing broadband X-ray imaging, spectroscopy, and timing. We processed the \nustar data with the standard Data Analysis Software (NuSTARDAS) v2.1.2, distributed in HEASoft v6.33.2 with the CALDB version 20240325. We used the statusexpr="STATUS==b0000xxx00xxxx000" keyword in the NuPIPELINE, as suggested for bright sources. The source and background are obtained by extracting the source in a circular region of 120\arcsec\ radius centered at the position of \gx and the background in an off-center sourceless circular region of 120\arcsec\ radius. The \nustar spectra were grouped to have a minimum of 30 counts per bin.
\end{appendix}
\end{document}